# Tunable Conductivity and Conduction Mechanism in a UV light activated electronic conductor

Mariana I. Bertoni[*], Thomas O. Mason[*], Julia E. Medvedeva[**], Arthur J. Freeman[**], Kenneth R. Poeppelmeier[†] and Bernard Delley[††]

[*] Department of Materials Science and Engineering, Northwestern University, Evanston, IL 60208, USA

[**] Department of Physics and Astronomy, Northwestern University, Evanston, IL 60208, USA

[†] Department of Chemistry, Northwestern University, Evanston, IL 60208, USA

[††] Paul Scherrer Institut, Villigen, CH-5232, Switzerland

## Abstract

A tunable conductivity has been achieved by controllable substitution of a novel UV light activated electronic conductor. The transparent conducting oxide system H-doped $Ca_{12-x}Mg_xAl_{14}O_{33}$ (x = 0; 0.1; 0.3; 0.5; 0.8; 1.0) presents a conductivity that is strongly dependent on the substitution level and temperature. Four-point dc-conductivity decreases with x from 0.26 S/cm (x = 0) to 0.106 S/cm (x = 1) at room temperature. At each composition the conductivity increases (reversibly with temperature) until a decomposition temperature is reached; above this value, the conductivity drops dramatically due to hydrogen recombination and loss. The observed conductivity behavior is consistent with the predictions of our first principles density functional calculations for the Mg-substituted system with x=0, 1 and 2. The Seebeck coefficient is essentially composition- and temperature-independent, the later suggesting the existence of an activated mobility associated with small polaron conduction. The optical gap measured remains constant near 2.6 eV while transparency increases with the substitution level, concomitant with a decrease in carrier content.

**Introduction**

Materials with low sheet resistance and optical transparency are sought for many optoelectronic applications. These two properties tend to be mutually exclusive in nature and are usually obtained by creating electron degeneracy in a wide band gap material [1]. Various transparent oxides are rendered electrically conducting by controllably introducing non-stoichiometry and/or appropriate dopants, but this approach is not applicable for the oxides of the main-group metals. Alternative processes are being developed to render these oxide systems conducting, as part of the search for inexpensive and environmentally benign alternatives [2].

The reported system, $12CaO \cdot 7Al_2O_3$ is a well-known insulating oxide widely used in high-alumina cements. The crystal lattice belongs to the cubic system and space group $I\bar{4}3d$ with a lattice parameter of 1.199 nm [3]. It possesses a cage structure with two formula units (12 cages) per unit cell and its empirical formula may be written as $[Ca_{24}Al_{28}O_{64}]^{4+} + 2O^{2-}$, where the free oxygen ions provide the charge neutrality and are located inside the cages of the framework (Fig. 1). The system was discovered to incorporate hydrogen at elevated temperatures through the following chemical reaction

$$O^{2-}_{(cage)} + H_{2(atmosphere)} \rightarrow OH^{-}_{(cage)} + H^{-}_{(cage)} \qquad (1)$$

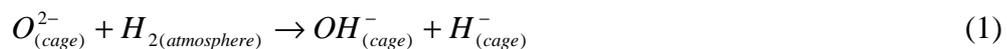

After hydrogen incorporation the unit cell contains two cages occupied by OH⁻, another two occupied by H⁻ and the remaining 8 cages of the unit cell are empty [2,4]. Hydrogen annealing results in no apparent change in the optical and electrical properties of the material. However, upon UV irradiation two optical absorption bands are induced, giving rise to a persistent color change from white to green together with a considerable conductivity increase. Our previous density functional investigations [4] revealed that the charge transport associated with the photo-excitation of an electron from hydrogen (which is located inside one of the cages, cf. Eq.(1)), occurs by electron variable-range hopping through states of the encaged $OH^-$ and $H^0$ [5] and d states of their nearest Ca neighbors (i.e., only 8 out of 24 Ca atoms in the unit cell of mayenite were found to give significant contributions to the density of states near the Fermi level [4].) The detailed knowledge of the transport mechanism thus obtained for H-doped UV-activated mayenite predicts the strong dependence of the light-induced conductivity on the atoms participating in the hopping as well as on their spatial arrangement and hence the possibility of varying the conductivity by proper doping. In particular, we expected that Mg substitution may lead to a decrease in the conductivity once Mg substitutes one of the Ca atoms involved in the hopping, since its 3d states will lie much higher in energy than those of Ca.

In this paper the conduction mechanism of the $Ca_{12-x}Mg_xAl_{14}O_{33}$ (x = 0; 0.1; 0.3; 0.5; 0.8; 1) system was studied both experimentally and theoretically. We report the change in conductivity, thermopower, optical gap (estimated from the absorption edge measured in a diffuse reflectance experiment) and transparency for the bulk oxide $Ca_{12-x}Mg_xAl_{14}O_{33}$. In addition, we present a small polaron conduction mechanism that is consistent with the electronic conduction in this system and discuss the conductivity behavior observed for $Ca_{12-x}Mg_xAl_{14}O_{33}$ based on the results and predictions of our first-principles total energy calculations for the Mg-substituted system with x=0, 1 and 2.

**Experimental**

Polycrystalline samples of Mg-substituted $12CaO \cdot 7Al_2O_3$ were produced by conventional high temperature solid-state reaction [6]. Stoichiometric amounts of $CaCO_3$, $Al_2O_3$ and MgO (>99.99% purity metals basis, Alfa Aesar) were mixed in an agate mortar to produce various samples of $Ca_{12-x}Mg_xAl_{14}O_{33}$ (x = 0; 0.1; 0.3; 0.5; 0.8; 1.0). Once homogeneous mixtures were obtained, 11.6 mm diameter x 2-3 mm thick pellets were pressed at 180 MPa. The pressed pellets were heated in air up to 1200 ºC in high-purity alumina crucibles and held for 24 h. Once cooled the pellets were reground, repressed and fired again at 1200 ºC for 24 h.

The phase purity of the samples was confirmed by powder x-ray diffraction (XRD) using CuKα radiation (Rigaku, Danvers, MA). A nickel filter was used to remove the CuK$_\beta$ contribution from the diffraction pattern. Powders were scanned between 10° and 80° in 2θ for routine phase analysis.

The subsequent hydrogen treatment was done inside an 18" length quartz tube sealed with a water-cooled metal end cap under a constant forming gas (4-5% H$_2$, N$_2$ balance) flow. The quartz tube was placed inside a high temperature tube furnace where the samples were taken to 1300 ºC for 2 hours and then rapidly cooled to room temperature by rapid extraction of the quartz tube out of the furnace.

After the hydrogenation, the samples were exposed to UV light under a mercury short arc lamp (Model HAS-200 D.C.) of total radiation (275 – 650 nm) 52.3 Watts, for 40-70 min.

Conductivity measurements were taken using the Van der Pauw technique where a 4-point spring-loaded probe touches the sample in four different points close to the edges [7]. A resistance $R_{AB,CD}$ is defined as the potential difference $V_D - V_C$ between the contacts D and C per unit current flowing through the contacts A and B. Similarly a resistance $R_{BC,DA}$ can be defined and the following relation holds for a specimen of arbitrary shape:

$$\exp\left(-\pi \cdot R_{AB,CD} \frac{d}{\rho}\right) + \exp\left(-\pi \cdot R_{BC,DA} \frac{d}{\rho}\right) = 1 \qquad (2)$$

where ρ is the resistivity of the material and d is the thickness of the UV activated slab. Corrections were made for layer thickness and sample diameter in every case. It should be mentioned that corrections for porosity were not performed due to the uncertainty in the properties of the irradiated layer. However, previous conductivity measurements on powder specimens using a powder solution composite technique (PSC) showed a very good agreement with the conductivity values obtained by the four point probe method [8].

Room temperature thermopower data were collected on bar-shaped samples cut from the irradiated pellets. The bars of approximately 10 mm x 3 mm x 3 mm had UV-activated conductive layers on one lateral face and both ends. These bar-shaped samples were sandwiched between two gold foil contacts. The bars were painted on the contact faces with a silver colloidal suspension to improve the electrical and thermal contact between the small UV-activated layer and the two gold electrodes. One gold contact was in thermal equilibrium with a 23 W heating element and the other was in thermal equilibrium with a cylindrical steel slug that rested on an insulating ceramic brick. A type S (Pt-Pt/10%Rh) thermocouple bead was welded to both gold contacts. A thermal gradient was created by switching on the heating element and allowing it to reach 100 ºC, at which point, the heating element was switched off, letting the system thermally relax. The temperature difference (ΔT) and the voltage difference (ΔV) were measured at regular intervals (3 sec) using a programmable scanner (Keithley 705, Cleveland, OH ) and a digital multimeter (Keithley 195A, Cleveland, OH) connected through an IEEE port to a personal computer. Thermopower was calculated by fitting the temperature and voltage gradient data with a least-squares fit as the sample approached equilibrium using the concept presented by Hong et al. [9]

$$Q = -\lim_{\Delta T \to 0} \frac{\Delta V}{\Delta T} \qquad (3)$$

A correction for the contribution of the Pt thermocouple to the overall thermopower has to be made using the polynomial fit of Hwang [10].

Since thin films of these materials were unavailable, optical data were obtained from diffuse reflectance measurements. The spectra for the specimens were collected on a Cary 500 UV-VIS-NIR spectrophotometer (Varian Instruments Inc., Palo Alto, CA) using a Diffuse Reflectance Accessory (DRA) between 400 and 700 nm with a lead sulfide detector. This accessory has the ability to collect most reflected radiation, remove any directional preferences, and present an integrated signal to the detector. The data were corrected with a spectrum obtained from a polytetrafluoroethylene (PTFE) reference specimen. The optical gap was estimated from the absorption edge that was determined by the intersection of a line drawn through the sloped portion of the transition region between high and low transmission and the baseline of the low-transmission portion of each spectrum.

**Theoretical**

First-principles all-electron density functional electronic structure calculations for pure and H-doped $Ca_{12-x}Mg_xAl_{14}O_{33}$ (x=0, 1 and 2) were performed using $DMol^3$ method [11]. The structures were modeled within the cell of mayenite with one formula unit (i.e., 59 atoms per cell which combine into six cages) with periodic boundary condition. The equilibrium relaxed geometry of the structures was determined via total energy and atomic forces minimization; during the optimization, all atoms were allowed to move in x, y and z directions, while the volume of the unit cell was fixed to the experimental value of the mayenite [3]. Summations over the Brillouin zone were carried out using 24 special k points in the irreducible wedge.

**Results and Discussion**

The temperature dependence of the conductivity for the system $Ca_{12-x}Mg_xAl_{14}O_{33}$ (x = 0; 0.1; 0.3; 0.5; 0.8; 1.0) is shown in Fig.2. Although the systematic error is on the order of 5% due to uncertainty in geometric factors, the random uncertainty is on the order of the symbol size or less. As shown in Fig. 2a, the conductivity is strongly dependent on the substitution level and rises reversibly from room temperature to approximately 130 ºC. Once the temperature exceeds 135 ºC (x = 0), the conductivity drops irreversibly as can be seen in Fig. 2b. This drop is related to the amount of hydrogen released when the system crosses the decomposition temperature and will be discussed in detail later. Figure 2a also shows a slight shift of the decomposition temperatures to higher values as the substitution level increases.

The shift of the decomposition temperature for the Mg-substituted mayenite can be explained based on the results of the density functional calculations for the H-doped $Ca_{11}MgAl_{14}O_{33}$. From a comparison of the total energies of the 12 fully relaxed structures with different Mg site locations, we found that (i) the Mg atom prefers to substitute for one of the Ca nearest neighbors of $H^-$, and (ii) the $H^-$ relaxes toward the Mg atom and demonstrates a strong Mg-H bonding, cf. Fig. 3. Consequently, more energy is required to release hydrogen from the sample as compared with H-doped UV-activated $Ca_{12}Al_{14}O_{33}$ – in agreement with our observations, cf. Fig. 2.

The thermoelectric coefficient was measured at room temperature for the different substitution levels (x = 0.1; 0.3; 0.5; 0.8; 1), and in the range between room temperature and 120 ºC for the pure $Ca_{12}Al_{14}O_{33}$ specimen. Results are shown in Fig. 4. The coefficient obtained was negative (n type) and, within experimental error, temperature- independent with a value of approximately –206 ± 12 µV/K. The fact that the conductivity is thermally activated while the Seebeck coefficient is temperature independent confirms that an activated mobility exists in the $Ca_{12-x}Mg_xAl_{14}O_{33}$ system; and this is indicative of a small polaron conduction mechanism [12].

The thermoelectric coefficient for small polaron conductors is given by [13]:

$$Q = \pm \frac{k}{e}\left(\ln\frac{2(1-c)}{c}\right) \qquad (4)$$

where k/e is 86.14 µV/K, c is the fraction of conducting ions of higher valence, the factor 2 accounts for the spin degeneracy, the entropy of the transport term is neglected, and the sign is determined by the nature of the polaron. It is negative if it forms around the trapped electron and positive for the trapped hole.

Based on the experimental result of the thermopower -206 µV/K ± 12 µV/K (see Fig. 2), the fraction of conducting species is constant and can be estimated as 0.155 ± 0.02, which is consistent with two carriers per unit cell moving along a 12-site hopping path (c = 2/12 = 0.166) as was suggested in our previous reports [4].

The expression for the small polaron conductivity is [14, 15]

$$\sigma = \frac{\sigma_o}{T}\exp\left(\frac{-E_H}{kT}\right) \qquad (5)$$

where $E_H$ is the activation or hopping energy and the pre-exponential factor $\sigma_o$ is given by

$$\sigma_o = \frac{gNc(1-c)e^2 a^2 \nu_o}{k} \qquad (6)$$

where g is a geometrical factor on the order of unity (related to the coordination number of equivalent sites), a is the jump distance between equivalent sites, $\nu_o$ is a lattice vibrational frequency, N is the total density of conducting sites, and c is the fraction of conducting species.

Figure 5a shows Arrhenius plots for the as-prepared and partially decomposed non-substituted sample (x = 0). It can be seen that there is no difference between the slopes of the graphs, confirming that the hopping energy, $E_h$, is the same for both cases and equal to 0.12 eV. This value compares favorably with values typical of small polaron behavior [16-18]. On the other hand the values of the pre-exponential factor are significantly different, showing a 35% decrease between the values before and after decomposition. Considering the values of a, g and $\nu_o$ in the pre-exponential expression (Eq. 6) to be approximately constant, and also recalling the constant value obtained for the fraction of occupied sites (c) from thermopower data, it can be concluded that the variation of the total density of conducting sites (N) is responsible for the 35% drop in the pre-exponential factor. To confirm this, Secondary Ion Mass Spectrometry (SIMS) measurements were performed on similarly treated deuterated samples. The amount of deuterium was detected before and after the decomposition temperature, giving a 38% loss that matches nicely with the 35 % drop in the total number of available sites. This result confirms not only that hydrogen loss occurs above a certain decomposition temperature, but also that the hydrogenous species are involved in the hopping path of the small polaron conduction model.

Similarly, Arrhenius plots for the Mg substituted samples are shown in Fig. 5b, where the $E_h$ is found to be the same for the different values of substitution (x = 0; 0.1; 0.3; 0.5; 0.8; 1.0) and equal to 0.12 eV. The pre-exponential factor obtained from the different intercepts of Fig. 5b drops dramatically with the substitution level and ranges from 16000 S K cm$^{-1}$ (x = 0) to 2370 S K cm$^{-1}$ (x = 1) as shown in Fig. 6.
Assuming again that g, a and $\nu_o$ have no significant change, and considering also that the value of c is approximately constant as calculated before from thermopower measurements, the different values obtained for $\sigma_o$ (see Fig. 6) have to be related to variations in the total density of conducting sites (N) with the substitution level. This suggests that once Mg substitutes on Ca sites in the cage structure of $Ca_{12}Al_{14}O_{33}$, the cages occupied by Mg are eliminated from the conduction process, as if Mg were a blocking agent.

This analysis can be taken one step further by considering the pre-exponential factor data normalized by that of the non-substituted sample (cf., Fig. 6). Figure 7 represents a complete unit cell (two formula units) showing only the atoms which contribute to the conduction mechanism. There are 12 conducting sites (8 Ca atoms, two $OH^-$ and two $H^0$) in 4 out of 12 cages that are involved in the conduction path. If magnesium ions have a tendency to occupy any of the calcium sites in the 4 conducting cages (as found by our total energy calculations mentioned above), and if each magnesium ion eliminates all of the 12 sites on the conducting path (i.e., it blocks the whole unit cell), we would expect an initial slope of -3 on Fig. 6, given by:

$$\frac{N_{(x)}}{N_{(x=0)}} = \frac{\sigma_o}{\sigma_{o(x=0)}} = 1 - \frac{12}{4}x = 1 - 3x \qquad (7)$$

As can be seen in Fig. 6, this slope is greater than the initial slope of the experimental values, suggesting that not all magnesium atoms are occupying conducting sites. Indeed, under the rapid cooling of the sample annealed at 1300 ºC, some of the Mg atoms can become "frozen" into the positions located far away from the hopping path although the corresponding total energy is found to be higher by at least 57 meV as compared to the most energetically favorable structure in which Mg atom substitutes one of the hopping centers of the UV-activated system.

Now, to further understand the conductivity behavior with an increase of the Mg concentration, we compared the total energy of 11 structures of $Ca_{10}Mg_2Al_{14}O_{33}$ with different site locations of one of the two Mg atoms (the other Mg atom was located at the most energetically favorable position as obtained from the calculations for $Ca_{11}MgAl_{14}O_{33}$). We found that the second Mg atom prefers to be located in the same cage with the first one [19], which thus demonstrates a tendency for Mg atoms to cluster.

Based on these results, we modify Eq. (7) by introducing the factor $\lambda$ which represents the fraction of magnesium in conducting cages. Moreover, if we consider that a second magnesium atom occupying a calcium site in a conducting cage results in no additional reduction of the conducting sites, the propensity of magnesium ions to cluster in the conducting cages can be accounted for by the term $\delta$ in the following equation:

$$\frac{N_{(x)}}{N_{(x=0)}} = \frac{\sigma_o}{\sigma_{o(x=0)}} = 1 - 3\lambda x(1-\delta x) \qquad (8)$$

The best fit to the experimental data in Fig. 6 is $\lambda=0.57$ (approximately 43% of the Mg ions occupying nonconducting cages) and $\delta=0.5$ (a significant tendency for Mg to cluster).

This small polaron model also allows to estimate the maximum conductivity achievable in the non-substituted mayenite sample, based on Eq. 5 and 6. We assume that the fraction of conducting sites occupied by carriers remains 2/12, a jump frequency of $\sim 10^{13}$ s$^{-1}$, and an average jump distance of 0.3 nm [4]. If we furthermore assume that all the O$^{2-}$ reacts with H$_2$ in the hydrogenation process, and that every H$^-$ releases an electron after UV irradiation, giving 4 conducting cages per unit cell, the total density of conducting sites will be 12 per unit cell. Using these values and setting the hopping energy to zero, Eq. 5 gives a maximum conductivity of $\sim 100$ S/cm, which agrees well with the results obtained by Matsuishi et al. on Ca-treated, fully reduced (no O$^{2-}$ species) mayenite [20].

Furthermore, the carrier content (Nc) can be estimated by combining the fraction of occupied sites (from thermopower) with the total density of conducting sites (from $\sigma_o$ coefficient). By combining the carrier content values with the conductivity, the mobility for the different substitution levels can be obtained as shown in Table I. These results are consistent with mobility values typical for small polaron conduction ($< 1$ cm$^2$/Vs) [21].

Finally, diffuse reflectance spectra shown in Fig. 8 characterize the optical properties of the $Ca_{12-x}Mg_xAl_{14}O_{33}$ system. Transmission increases with the doping level; however, the energy of the optical gap remains constant at 2.6 eV and is in good agreement with previous results [2]. Increased transmission is consistent with the drop in the number of carriers from the electrical measurements (see Table I).

**Conclusions**

Electrical property measurements of $Ca_{12-x}Mg_xAl_{14}O_{33}$ (x = 0; 0.1; 0.3; 0.5; 0.8; 1.0) offer evidence for small polaron conduction. The electrical conductivity is activated, whereas the thermoelectric coefficient remains temperature-independent. The activation energy obtained for all the compositions is the same and equal to 0.12 eV. Carrier contents on the order of $10^{20} - 10^{21}$ cm$^{-3}$ were calculated from the conductivity pre-exponential factors and the thermopower values (~ -206 μV/K). The mobilities obtained Mg-substituted mayenite at 125 °C range from 4 x $10^{-3}$ to 9 x $10^{-3}$ cm$^{-2}$/Vs and are consistent with the small polaron transport mechanism. This study shows that hydrogen is intimately involved in the conduction process, and loss of hydrogen above a decomposition temperature leads to a permanent reduction in conductivity. The optical absorption peak induced by the irradiation is 2.6 eV and does not shift with Mg concentration. For wavelengths longer than 500 nm there is a monotonic decrease in adsorption as the substitution level increases, which is consistent with the drop in electron population.

Finally, we found, both theoretically and experimentally, that Mg in the $Ca_{12}Al_{14}O_{33}$ system acts as a blocking agent on the conduction path. Although this produces a compositionally tunable conductivity, the resulting conductivities are limited by the low mobilities associated with small polaron conduction.

**Acknowledgments**

This work was supported by the MRSEC program of the National Science Foundation (DMR-0076097) at the Materials Research Center at Northwestern University and by the DOE (DE-FG02-88ER45372). M.I.B is supported by the U. S. Department of State through a Fulbright Scholarship. Computational resources have been provided by the DOE supported NERSC.

**References**

[1] B. Lewis, and D. Paine, MRS Bull. **25**(8), 22 (2002).

[2] K. Hayashi, S. Matsuichi, T. Kamiya, M. Hirano, and H. Hosono, Nature (London) **419**, 462 (2002).

[3] V. Bartl, and T. Scheller, T. Neues Jahrb., Mineral., Monatsh. **35**, 547 (1970); A.N. Christensen, Acta Chem. Scand. Ser. A **42**, 110 (1987).

[4] J. E. Medvedeva, A. J. Freeman, M. I. Bertoni, and T. O. Mason, Phys. Rev. Lett. **93**, 016408 (2004): J. E. Medvedeva, and A. J. Freeman, Appl. Phys. Lett. **85**, 955 (2004).

[5] There is some uncertainty regarding the precise UV-activated mechanism. An alternative involves the reaction: $H^- \rightarrow H^+ + 2e^-$ and is the subject of ongoing work.

[6] H. Hosono, and Y. Abe, Inorg. Chem **26**, 1192 (1987).

[7] L. Van der Pauw, Philips Res. Repts. **13**, 1 (1958).

[8] B. Ingram, and T. Mason, J. Electrochem. Soc. **150**(8), E396 (2003).

[9] B. Hong, S. Ford, and T. Mason, Key Eng. Mat. **125-126**, 163 (1997).

[10] J. Hwang, PhD Thesis, Northwestern University, Evanston, Illinois (1996).

[11] B. Delley, J. Chem. Phys. **113**, 7756 (2000).

[12] J. Nell, B. J. Wood, S. E. Dorris, and T. O. Mason, J. Sol. State Chem. **82**, 247 (1989).

[13] A. F. Joffe, Physics of Semiconductors. Infosearch Ltd., London (1960).

[14] H. L. Tuller, and A. S. Nowick, J. Phys. Chem. Solids **38**, 859 (1977).

[15] R. Dieckmann, C. A. Witt, and T. O. Mason, Ber. Bunsen-Ges. Phys. Chem. **87**(6), 495 (1983).

[16] D. P. Karim, and A. T. Aldred, Phys. Rev. **B20**, 2255 (1979).

[17] T. O. Mason, and H. K. Bowen, J. Amer. Ceram. Soc. **64**, 237 (1981).
- 14 -

Table I. Transport data for $Ca_{12-x}Mg_xAl_{14}O_{33}$ at room temperature and 125 °C

| Substitution Level (x) | Carrier Content ($cm^{-3}$) | Conductivity 25°C (S/cm) | Mobility 25°C ($cm^{-2}/Vs$) | Conductivity 125°C (S/cm) | Mobility 125°C ($cm^{-2}/Vs$) |
|---|---|---|---|---|---|
| 0.0 | 1.08E+21 | 0.270 | 1.57E-03 | 0.711 | 4.12E-03 |
| 0.1 | 6.93E+20 | 0.242 | 2.18E-03 | 0.534 | 4.81E-03 |
| 0.3 | 6.49E+20 | 0.209 | 2.01E-03 | 0.515 | 4.95E-03 |
| 0.5 | 3.59E+20 | 0.165 | 2.86E-03 | 0.384 | 6.67E-03 |
| 0.8 | 2.42E+20 | 0.131 | 3.37E-03 | 0.318 | 8.18E-03 |
| 1.0 | 1.58E+20 | 0.106 | 4.20E-03 | 0.234 | 9.26E-03 |

Fig. 1. One of 12 cages constituting the unit cell of $Ca_{12}Al_{14}O_{33}$. The charge neutrality is given by the $O^{2-}$ located inside the cage.

Fig. 2. (a) Conductivity vs temperature of $Ca_{12-x}Mg_xAl_{14}O_{33}$ (x= 0, 0.1, 0.3, 0.5, 0.8, 1), (b) Conductivity vs. temperature of pure mayenite ($Ca_{12}Al_{14}O_{33}$) showing irreversible loss of conductivity above the decomposition temperature (135 °C).

Fig. 3. Calculated total energy, in eV, as a function of the $H^-$ location between Mg and Ca atoms in H-doped $Ca_{11}MgAl_{14}O_{33}$. To obtain this curve, the positions of all atoms in the cell except $H^-$ were fixed. In the insert: The calculated distances, in Angstroms, between the $H^-$ ion and its nearest Ca (Mg) neighbors in the fully relaxed (i.e., all atoms in the cell were allowed to move) structures of (a) $Ca_{12}Al_{14}O_{33}$ and (b) $Ca_{11}MgAl_{14}O_{33}$ with the Mg atom located in the most energetically favorable position.

Fig. 4. Thermopower vs. substitution level at room temperature and 105 °C.

Fig. 5. (a) Arrhenius plot of electrical properties for as-prepared and partially-decomposed non-substituted mayenite (Ca12Al14O33) from the data in Fig. 2b. (b) Arrhenius plot of electrical properties for Ca12-xMgxAl14O33 (x = 0, 0.1, 0.3, 0.5, 1)

Fig. 6. Pre-exponential factor and normalized pre-exponential factor vs. substitution level.

Fig. 7. Hopping path for the conduction mechanism of $Ca_{12}Al_{14}O_{33}$

Fig 8. Optical properties for $Ca_{12-x}Mg_xAl_{14}O_{33}$

M. Bertoni, Fig 1

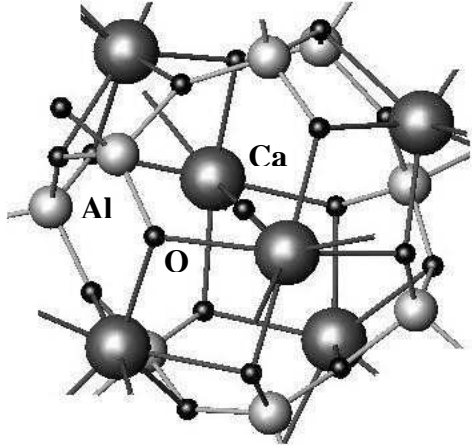

M. Bertoni, Fig 2

(a)

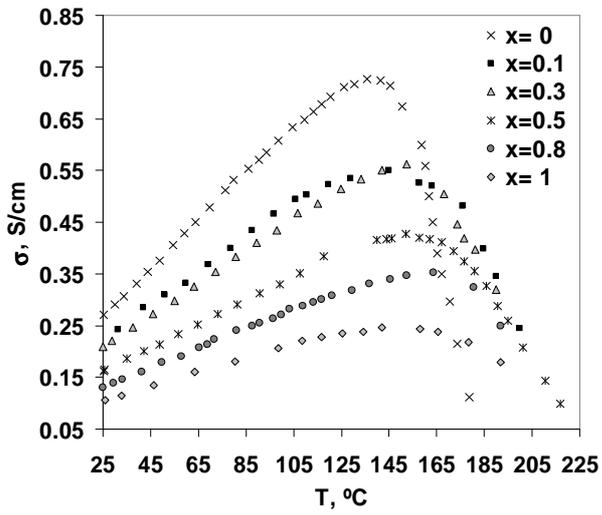

(b)

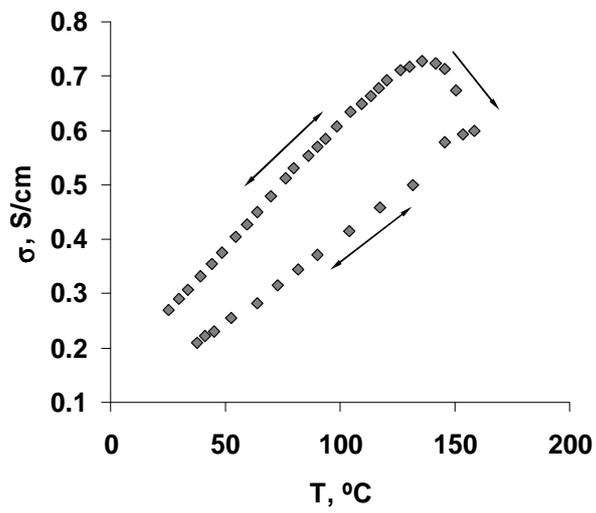

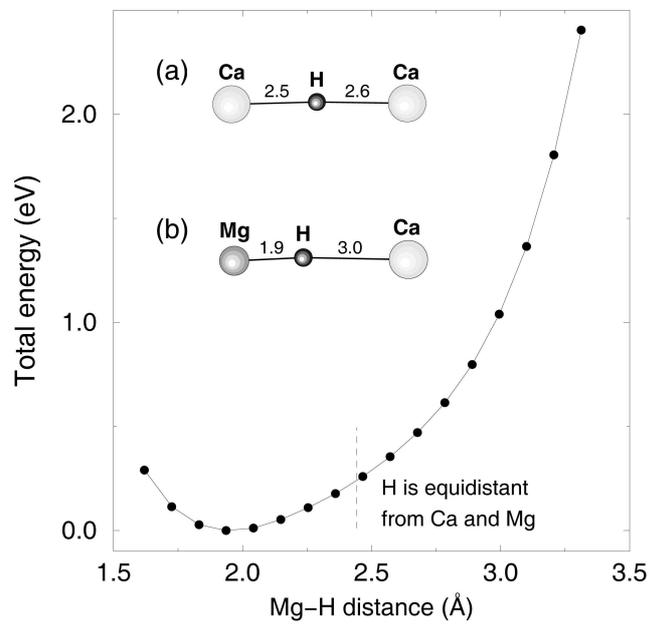

M. Bertoni, Fig 4

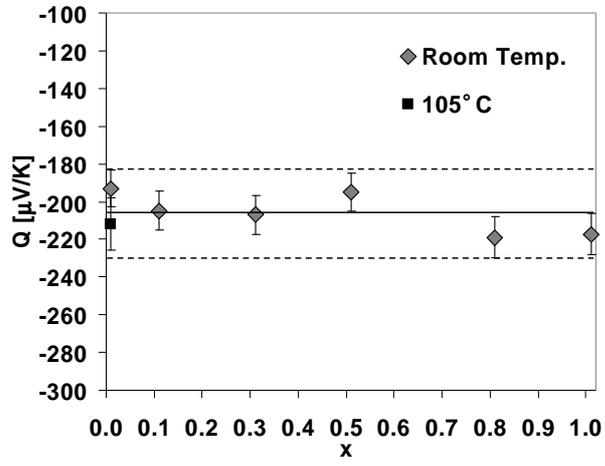

M. Bertoni, Fig 5.

(a)

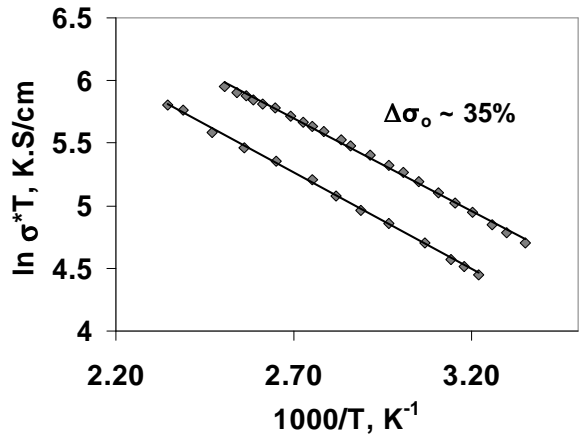

(b)

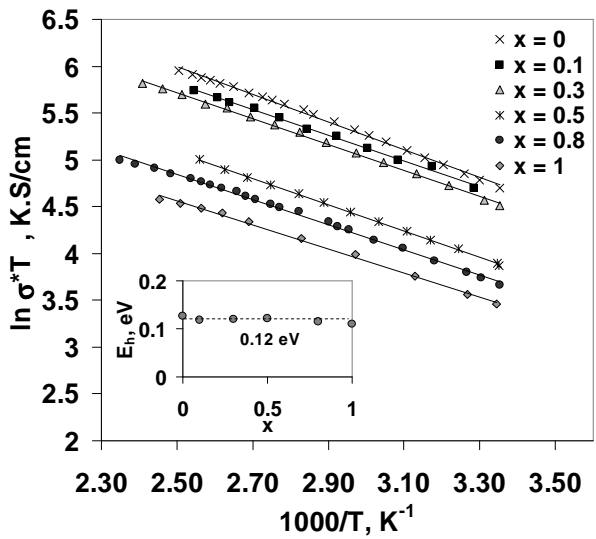

M. Bertoni, Fig 6.

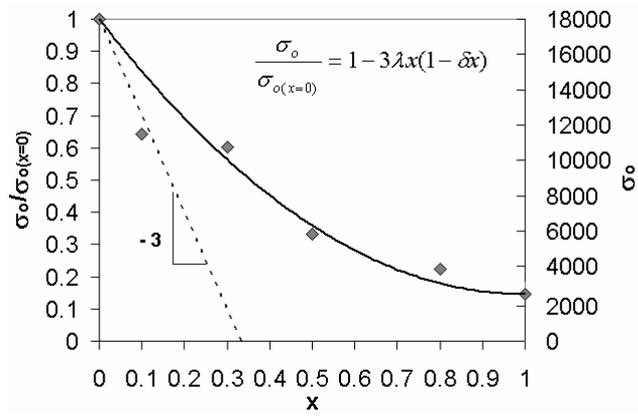

M. Bertoni, Fig 7.

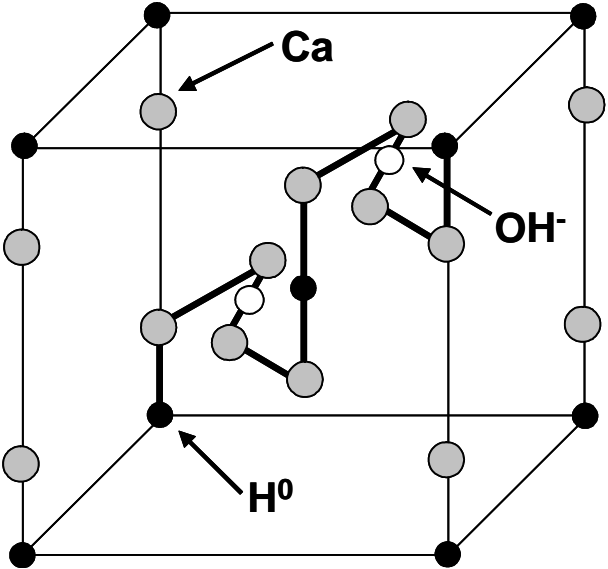

M. Bertoni, Fig 8.

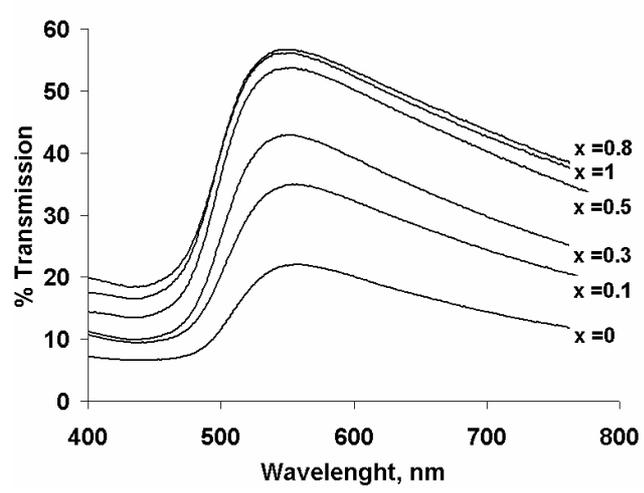